\documentclass[lettersize,journal]{IEEEtran}

\usepackage{amsmath,amssymb,amsfonts}
\usepackage{graphicx}
\usepackage{cite}
\usepackage{array}
\usepackage{booktabs}
\usepackage{algorithm}
\usepackage{algpseudocode}
\usepackage{amsmath}
\usepackage[table]{xcolor}
\usepackage{enumitem}

\begin{document}

\title{QSVT-Based Three-Phase Unbalanced Power Flow} 

\author{
    Kamini~Shahare and Peng~Zhang

\thanks{This material is based upon work supported by the National Science Foundation under Grant Number 2432387.}
    
    \thanks{
    K. Shahare and P. Zhang are with the Department of Electrical and Computer Engineering,
    Stony Brook University, Stony Brook, NY 11794-2350, USA
    (e-mail: kamini.shahare,  p.zhang@stonybrook.edu).
    }
}

\maketitle

\begin{abstract}
This letter introduces QSVT-3PF, a quantum singular value transformation (QSVT) based solver for three-phase unbalanced power flow with embedded single-phase grid-forming inverter  (GFM) operation. The contributions are threefold: 1) reformulating the Newton correction as a QSVT-compatible inverse problem using a normalized block-encoded phase-domain Jacobian; 2) introducing a regularized singular-value filter to improve robustness under ill-conditioned and stressed operating conditions; and 3) validating the proposed solver on an IEEE 5-bus system and the IEEE 123-node test feeder with single-phase GFM integration. Test results show that QSVT-3PF 
matches the classical Newton benchmark in residual convergence, voltage profile, and final operating point, demonstrating the feasibility of QSVT-3 for large unbalanced distribution systems.
\end{abstract}



\section{Introduction}

As distribution networks become larger, more unbalanced, and more inverter-dominated, scalable alternatives to classical three-phase power flow 
are increasingly attractive. Quantum computing provides a promising pathway to accelerate grid analytics such as power flow~\cite{b1,b2}.  
Existing quantum power flow studies have primarily considered HHL-type methods or variational quantum linear solvers~\cite{b3}. However, HHL requires demanding assumptions on state preparation and circuit depth, while variational methods depend on classical optimization and may suffer from trainability limitations.

Quantum singular value transformation (QSVT) provides a general framework for implementing matrix functions through polynomial transformations of the singular values of a block-encoded matrix~\cite{b4}. Since Newton-based power flow repeatedly requires the inverse action of the Jacobian on the residual vector, QSVT is well suited for developing a quantum-compatible power flow solver.

This letter proposes QSVT-3PF, a quantum singular value transformation-based solver for three-phase unbalanced power flow with embedded single-phase grid-forming inverter operation. The main contributions are summarized as follows:
\begin{enumerate}[leftmargin=*] 
\item We reformulate the Newton correction for three-phase power flow as a QSVT-compatible matrix-inversion problem with a normalized block-encoded Jacobian.

\item We develop a polynomial inverse singular-value filter with regularized QSVT updates to compute robust Newton corrections without variational optimization.

\item We validate the proposed QSVT-3PF solver on the IEEE 5-bus system and IEEE 123-node test feeder with single-phase GFM integration, demonstrating its applicability to unbalanced inverter-dominated distribution networks.

\end{enumerate}

\section{Three-Phase Unbalanced Power Flow Problem}

The underlying formulation follows the single-phase GFM-embedded three-phase power-flow framework in~\cite{b5}. A single-phase GFM is connected to of the bus in the system called GFM bus. It is represented through its phase-specific network injection and active-power/frequency droop behavior. Hence, the conventional three-phase residual is augmented with a droop-frequency residual, and the common frequency $\omega$ is solved together with the phase-domain voltages.

Let $\mathcal{P}$ denote the energized bus phases and let $\mathbf{V}_{abc}\in\mathbb{C}^{|\mathcal{P}|}$ be the phase-domain voltage vector. The real-valued augmented residual is
\begin{equation}
    \mathbf{r}(\mathbf{x})
    =
    \begin{bmatrix}
    \mathbf{P}_{\mathrm{spec}}-\Re\{\mathbf{S}(\mathbf{V}_{abc})\}\\
    \mathbf{Q}_{\mathrm{spec}}-\Im\{\mathbf{S}(\mathbf{V}_{abc})\}\\
    \Delta W
    \end{bmatrix},
    \qquad
    \mathbf{x}
    =
    \begin{bmatrix}
    \Re\{\mathbf{V}_{abc}\}\\
    \Im\{\mathbf{V}_{abc}\}\\
    \omega
    \end{bmatrix}
\end{equation}
where $\Delta W$ denotes the GFM droop-frequency residual.

At each Newton iteration, the correction satisfies
$\mathbf{J}(\mathbf{x}^{k})\Delta\mathbf{x}^{k}=-\mathbf{r}(\mathbf{x}^{k})$.
The inverse action of this augmented three-phase Jacobian is the computational bottleneck and is to be replaced by the proposed QSVT-based singular-value filtering step as follows.

\section{QSVT-Based Newton Correction}

The key objective of QSVT-3PF is to approximate the Newton correction
using quantum singular value transformation. \vspace{-10pt}
\subsection{Jacobian Normalization}

The Jacobian is first normalized as
\begin{equation}
    \widetilde{\mathbf{J}}
    =
    \frac{\mathbf{J}}{\alpha}
\end{equation}
where
\begin{equation}
    \alpha \geq \|\mathbf{J}\|_2 
\end{equation}
This normalization guarantees that the singular values of $\widetilde{\mathbf{J}}$ lie within $[0,1]$.

Let the singular value decomposition of $\widetilde{\mathbf{J}}$ be
\begin{equation}
    \widetilde{\mathbf{J}}
    =
    \mathbf{U}
    \boldsymbol{\Sigma}
    \mathbf{V}^{H}
\end{equation}
where
\begin{equation}
    \boldsymbol{\Sigma}
    =
    \mathrm{diag}
    \left(
    \sigma_1,\sigma_2,\ldots,\sigma_n
    \right)
\end{equation}

The exact inverse action can be written as
\begin{equation}
    \mathbf{J}^{-1}
    =
    \frac{1}{\alpha}
    \mathbf{V}
    \boldsymbol{\Sigma}^{-1}
    \mathbf{U}^{H}
\end{equation}

Therefore, the Newton correction becomes
\begin{equation}
    \Delta \mathbf{x}
    =
    -
    \frac{1}{\alpha}
    \mathbf{V}
    \boldsymbol{\Sigma}^{-1}
    \mathbf{U}^{H}
    \mathbf{r}
\end{equation}

\subsection{QSVT Inverse Polynomial}

QSVT implements the desired singular-value transformation through a polynomial approximation. For Newton power flow, the target operation is the inverse singular-value map, $f(\sigma)=1/\sigma$. Since direct inversion becomes ill-conditioned near zero, the approximation is restricted to the well-conditioned interval and represented as
\begin{equation}
    p_d(\sigma)\approx \frac{1}{\sigma},
    \qquad
    \sigma \in [\sigma_{\min},1],
    \quad \sigma_{\min}>0 .
\end{equation}

Using this polynomial filter, the inverse action of the normalized Jacobian is approximated as
\begin{equation}
    \mathbf{J}^{-1}
    \approx
    \frac{1}{\alpha}
    \mathbf{V}
    p_d(\boldsymbol{\Sigma})
    \mathbf{U}^{H}.
\end{equation}
Thus, the QSVT-based Newton correction is
\begin{equation}
    \Delta \mathbf{x}_{\mathrm{QSVT}}
    =
    -
    \frac{1}{\alpha}
    \mathbf{V}
    p_d(\boldsymbol{\Sigma})
    \mathbf{U}^{H}
    \mathbf{r}.
\end{equation}

\subsection{Regularized QSVT Filter}
For ill-conditioned cases, a regularized inverse filter is used to suppress small singular-value amplification:
\begin{equation}
    p_{d,\lambda}(\sigma)
    \approx
    \frac{\sigma}{\sigma^2+\lambda},
    \qquad \lambda>0 
\end{equation}
The regularized QSVT Newton correction is
\begin{equation}
    \Delta \mathbf{x}_{\mathrm{QSVT}}
    =
    -
    \frac{1}{\alpha}
    \mathbf{V}
    p_{d,\lambda}(\boldsymbol{\Sigma})
    \mathbf{U}^{H}
    \mathbf{r}
\end{equation}

\section{Quantum Block-Encoding Formulation}

To implement QSVT, the normalized Jacobian $\widetilde{\mathbf{J}}$ is embedded into a larger unitary operator. A block encoding of $\widetilde{\mathbf{J}}$ is defined as a unitary $\mathbf{U}_{J}$ satisfying
\begin{equation}
    \left(
    \langle 0|^{\otimes a}
    \otimes
    \mathbf{I}
    \right)
    \mathbf{U}_{J}
    \left(
    |0\rangle^{\otimes a}
    \otimes
    \mathbf{I}
    \right)
    =
    \widetilde{\mathbf{J}}
\end{equation}
where $a$ is the number of ancilla qubits.

Given this block encoding, QSVT constructs a sequence of alternating applications of $\mathbf{U}_{J}$, $\mathbf{U}_{J}^{\dagger}$, and single-qubit phase rotations:
\begin{equation}
    \mathbf{U}_{\Phi}
    =
    e^{j\phi_0 Z}
    \prod_{\ell=1}^{d}
    \left(
    \mathbf{U}_{J}
    e^{j\phi_{\ell} Z}
    \mathbf{U}_{J}^{\dagger}
    e^{j\phi_{\ell}' Z}
    \right)
\end{equation}
where phase angles $\Phi=\{\phi_{\ell}\}$ are chosen so that the resulting SVT implements the polynomial $p_d(\sigma)$.

The effective action of the QSVT circuit is
\begin{equation}
    \widetilde{\mathbf{J}}
    \longrightarrow
    p_d(\widetilde{\mathbf{J}})
\end{equation}

When applied to the 
encoding of the residual vector,
\begin{equation}
    |\mathbf{r}\rangle
    =
    \frac{\mathbf{r}}{\|\mathbf{r}\|}
\end{equation}
the QSVT circuit produces a state proportional to
\begin{equation}
    p_d(\widetilde{\mathbf{J}})
    |\mathbf{r}\rangle
\end{equation}

After rescaling by $\|\mathbf{r}\|/\alpha$, the resulting vector approximates the Newton correction:
\begin{equation}
    \Delta \mathbf{x}_{\mathrm{QSVT}}
    \approx
    -
    \frac{\|\mathbf{r}\|}{\alpha}
    p_d(\widetilde{\mathbf{J}})
    |\mathbf{r}\rangle
\end{equation}

Algorithm 1 provides the pseudo code for QSVT-3PF. 


\begin{algorithm}[t]
\caption{QSVT-3PF Solver}
\label{alg:qsvt_3pf}
\begin{algorithmic}[1]
\State \textbf{Input:} $\mathbf{Y}_{abc}$, load data, GFM droop, $\mathbf{V}_{abc}^{0}$, $\epsilon$, $k_{\max}$
\State \textbf{Output:} $\mathbf{V}_{abc}$ and residual history
\State Initialize $\mathbf{x}^{0}=[\Re\{\mathbf{V}_{abc}^{0}\}^{T},\Im\{\mathbf{V}_{abc}^{0}\}^{T}]^{T}$

\For{$k=0,1,\ldots,k_{\max}$}
    \State Compute $\mathbf{I}_{abc}^{k}$, $\mathbf{S}_{abc}^{k}$, and residual $\mathbf{r}^{k}$
    \If{$\|\mathbf{r}^{k}\|_{\infty}<\epsilon$}
        \State \textbf{break}
    \EndIf

   \State Form the three-phase Jacobian $\mathbf{J}^{k}$
    \State Normalize $\widetilde{\mathbf{J}}^{k}=\mathbf{J}^{k}/\alpha^{k}$, $\alpha^{k}\geq\|\mathbf{J}^{k}\|_{2}$
    \State Apply either the inverse filter $p_d(\sigma)\approx 1/\sigma$ or the regularized filter $p_{d,\lambda}(\sigma)\approx \sigma/(\sigma^2+\lambda)$
    \State Compute the QSVT Newton correction:
    \[
    \Delta\mathbf{x}_{q}^{k}
    =
    -
    \frac{1}{\alpha^{k}}
    p(\widetilde{\mathbf{J}}^{k})\mathbf{r}^{k}
    \]
    \State Update $\mathbf{x}^{k+1}=\mathbf{x}^{k}+\Delta\mathbf{x}_{q}^{k}$
    \State Recover $\mathbf{V}_{abc}^{k+1}$ from $\mathbf{x}^{k+1}$
\EndFor

\State \Return $\mathbf{V}_{abc}$ and $\{\|\mathbf{r}^{k}\|_{\infty}\}$
\end{algorithmic}
\end{algorithm}

\section{Test Case Scenarios}

We validate QSVT-3PF on two test systems: an IEEE 5-bus system and the IEEE 123-node test feeder. The IEEE 5-bus system is used to verify the QSVT inverse approximation in a compact network, while the IEEE 123-node feeder is used to evaluate the method under feeder-scale three-phase unbalanced conditions with single phase GFM.

\subsection{IEEE 5-Bus System}

The IEEE 5-bus system is used as a compact benchmark to verify the basic performance of the proposed QSVT-3PF solver before applying it to larger unbalanced feeders. The system is modeled in three-phase coordinates with a single-phase GFM inverter connected at Bus~5, Phase~A. This small case allows the iteration-wise behavior of the QSVT-based inverse approximation to be clearly observed and directly compared with the classical Newton solution.

\begin{figure}[t]
    \centering
    \includegraphics[width=0.48\textwidth]{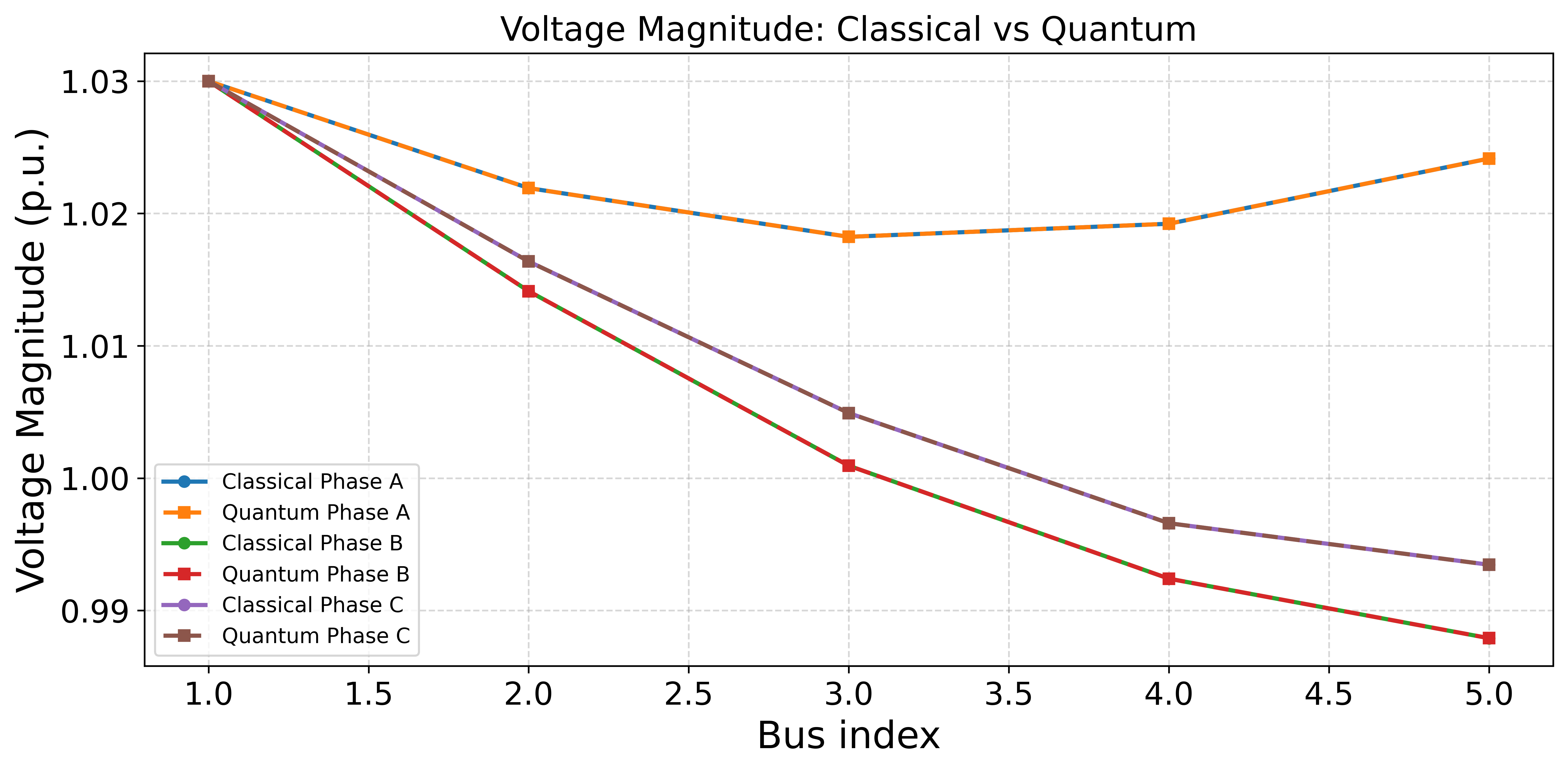}
    \caption{IEEE 5-bus system: convergence and voltage magnitude comparison between classical Newton and QSVT-3PF.}
    \label{fig:ieee5}
\end{figure}

As shown in Fig.~\ref{fig:ieee5}, the Classical Newton PF and Quantum QSVT PF exhibit nearly identical convergence behavior. Both solvers reach the same final voltage profile and droop-regulated frequency, confirming that the QSVT singular-value filtering step produces an equivalent Newton correction for the augmented three-phase Jacobian. The close overlap of the voltage magnitudes further verifies that the QSVT-3PF formulation preserves the classical steady-state operating point while replacing the direct linear solve with a polynomial inverse approximation.


\subsection{IEEE 123-Node Test Feeder}

\begin{figure}[h]
    \centering
    \includegraphics[width=0.48\textwidth]{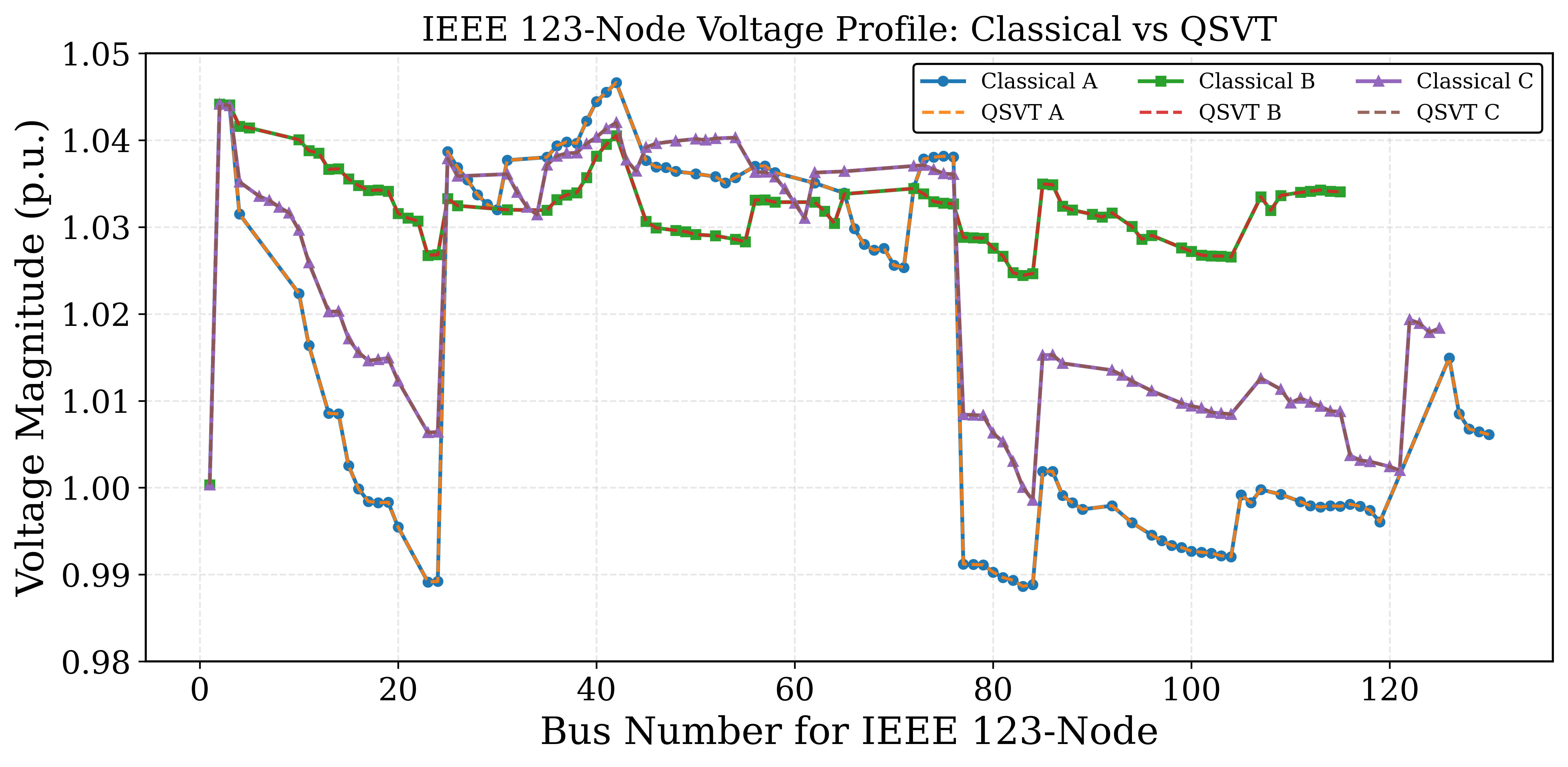}
    \caption{IEEE 123-node test feeder: voltage magnitude profile obtained using the proposed QSVT-3PF solver.}
    \label{fig2}
\end{figure}

The IEEE 123-node test feeder is used as the large-scale unbalanced benchmark to evaluate the scalability of the proposed QSVT-3PF solver. Compared with the IEEE 5-bus case, this feeder includes a larger number of voltage nodes, phase-specific laterals, and highly unbalanced loading, making it suitable for feeder-scale validation. Fig.~\ref{fig2} shows the feeder-wide voltage magnitude profile obtained using the proposed QSVT-3PF solver for the IEEE 123-node test feeder.

Table~\ref{tab:qsvt_newton_iter} presents the iteration-wise comparison of Quantum QSVT PF and Classical Newton PF for three representative GFM-connected voltage quantities, $V_{40A}$, $V_{82B}$, and $V_{114A}$, along with the frequency variable $\omega$. These monitored quantities are selected from different feeder locations and phases to evaluate whether the QSVT inverse approximation preserves the classical Newton update across the unbalanced network. Both solvers produce identical values at every iteration and converge at the fourth iteration to $V_{40A}=0.9979$ pu, $V_{82B}=1.0395$ pu, $V_{114A}=1.0254$ pu, and $\omega=0.9998$ pu. The identical trajectories verify that the QSVT singular-value filter accurately reproduces the Newton correction and preserves the final operating point of the IEEE 123-node feeder.

\begin{table}[h]
\centering
\caption{Iteration-wise comparison of Quantum QSVT PF and Classical Newton PF}
\label{tab:qsvt_newton_iter}
\scriptsize
\renewcommand{\arraystretch}{1.15}
\setlength{\tabcolsep}{4pt}
\resizebox{\columnwidth}{!}{
\begin{tabular}{c c c c c c}
\hline
\hline
\textbf{Algorithm} & \textbf{Iteration} & $\mathbf{V}_{40A}$ & $\mathbf{V}_{82B}$ & $\mathbf{V}_{114A}$ & $\boldsymbol{\omega}$ \\
\hline
Quantum QSVT PF & 1  & 0.9597 & 1.0001 & 0.9858 & 0.9985  \\
                & 2  & 0.9987 & 1.0403 & 1.0261 & 0.9998  \\
                & 3  & 0.9979 & 1.0395 & 1.0254 & 0.9998  \\
\rowcolor{blue!15}
                & 4$^{*}$ & 0.9979 & 1.0395 & 1.0254 & 0.9998  \\
\hline
Classical Newton PF & 1  & 0.9597 & 1.0001 & 0.9858 & 0.9985  \\
                    & 2  & 0.9987 & 1.0403 & 1.0261 & 0.9998  \\
                    & 3  & 0.9979 & 1.0395 & 1.0254 & 0.9998  \\
\rowcolor{blue!15}
                    & 4$^{*}$ & 0.9979 & 1.0395 & 1.0254 & 0.9998  \\
\hline
\hline
\end{tabular}
}
\end{table}

\begin{figure}[h]
    \centering
    \includegraphics[width=0.48\textwidth]{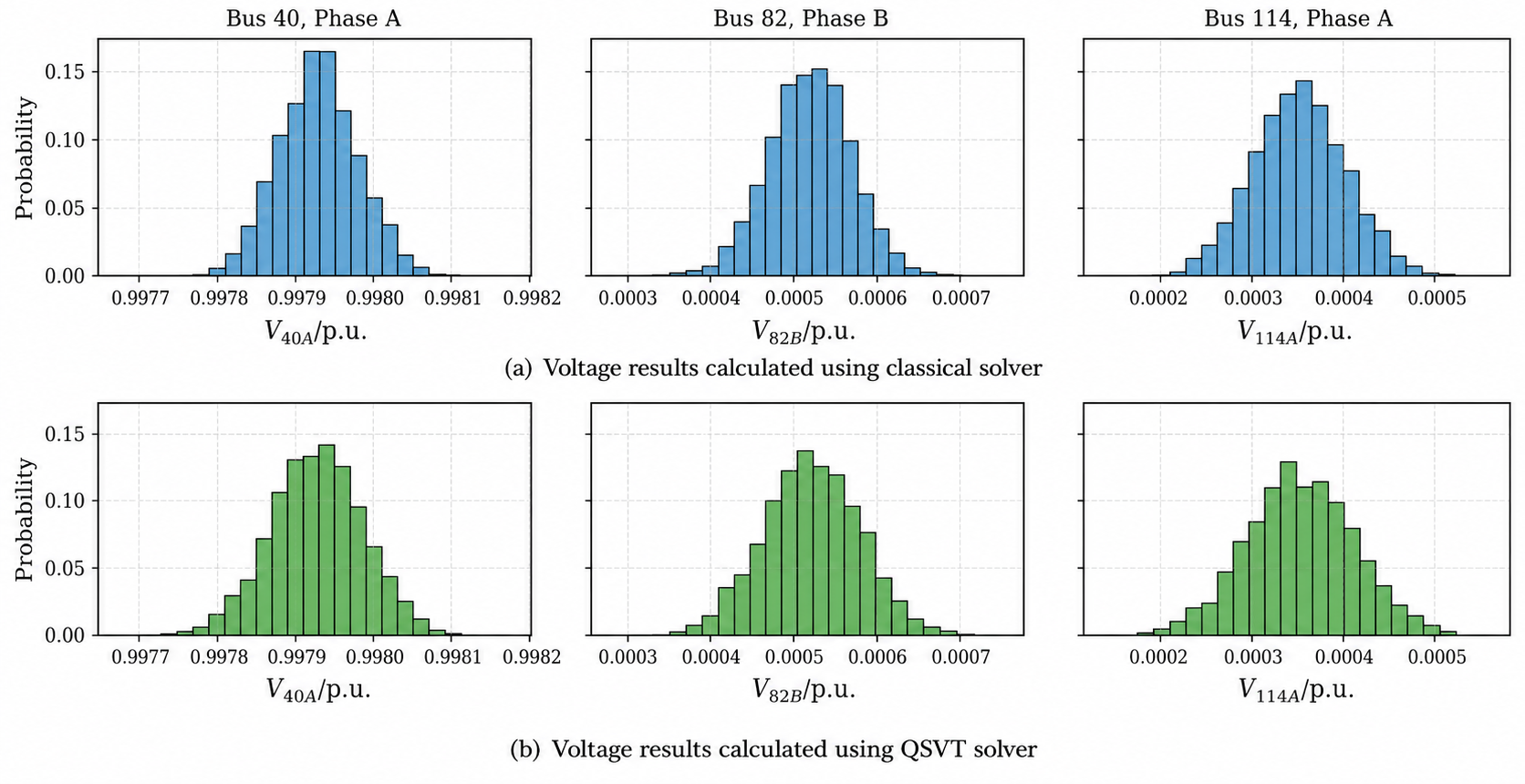}
    \caption{Probability distribution of voltage magnitudes at GFM 
    buses. }
    \label{fig3}
\end{figure}

Fig.~\ref{fig3} presents the probability distributions of voltage magnitudes at the GFM-connected buses. These histograms provide a statistical view of the voltage concentration around the converged operating point. The close agreement between the classical Newton and QSVT-based distributions indicates that the QSVT inverse approximation preserves the voltage solution at the GFM buses and does not introduce noticeable numerical dispersion. Since the QSVT solution produces nearly the same concentration pattern as the classical Newton solution, the dominant voltage states are preserved after the singular-value filtering process. This confirms that the reduced quantum inverse-filter step remains consistent with the full Newton-based feeder solution.


\section{Conclusion}

This letter demonstrates that three-phase unbalanced power flow
can be recast from a direct Newton solve into a singular-value
filtering problem. In practice, the network equations and the three-phase Jacobian
are formed in the same way as in a classical Newton power-flow
program; only the Jacobian-solve step used to compute the voltage
correction is replaced by the QSVT inverse filter. The new  QSVT-3PF formulation therefore provides a 
quantum path to preserve Newton-like accuracy while enabling
spectral regularization of ill-conditioned operating points.  Particularly, its regulation of weak singular directions of the Jacobian makes it well suited for stressed,
unbalanced, and inverter-rich feeders.

In sum,
QSVT-3PF offers a reusable computational primitive for repeated
grid analytics, such as feasible operating region analysis, DER hosting
studies, and future quantum-enabled digital twins for distribution grids.

\bibliographystyle{IEEEtran}

\end{document}